\documentclass[conference]{IEEEtran}

\usepackage[
    utf8
]{inputenc}

\usepackage[
  anchorcolor=blue,
  colorlinks=true,
  linkcolor=blue,
  urlcolor=black,
  citecolor=blue
]{hyperref}

\usepackage[
  xcdraw,
  table
]{xcolor}

\usepackage{amsmath}
\usepackage{amssymb}
\usepackage{graphicx}
\usepackage{multirow}
\usepackage{subfig}

\makeatletter
\g@addto@macro\normalsize{
    \setlength\belowdisplayskip{5pt}
    \setlength\abovedisplayskip{5pt}
}
\makeatother

\begin{document}

\title{Complex Network Tools to Understand the Behavior of Criminality in Urban Areas}

\author{
    Gabriel Spadon, Lucas C. Scabora, Marcus V. S. Araujo, Paulo H. Oliveira,\\
    Bruno B. Machado, Elaine P. M. Sousa, Caetano Traina-Jr, Jose F. Rodrigues-Jr\\
    
    University of Sao Paulo, Brazil\\
    
    \{spadon, lucascsb, pholiveira, araujo\}@usp.br
    ~
    \{brandoli, parros, caetano, junio\}@icmc.usp.br
}

% make the title area
\maketitle

\begin{abstract}
    % Context
    Complex networks are nowadays employed in several applications.
    Modeling urban street networks is one of them, and in particular to analyze criminal aspects of a city. 
    % Gap
    Several research groups have focused on such application, but until now, there is a lack of a well-defined methodology for employing complex networks in a whole crime analysis process, \textit{i.e.} from data preparation to a deep analysis of criminal communities. 
    Furthermore, the ``toolset'' available for those works is not complete enough, also lacking techniques to maintain up-to-date, complete crime datasets and proper assessment measures.
    % Purpose
    In this sense, we propose a threefold methodology for employing complex networks in the detection of highly criminal areas within a city.
    % Methodology
    Our methodology comprises three tasks: (i) Mapping of Urban Crimes; (ii) Criminal Community Identification; and (iii) Crime Analysis.
    Moreover, it provides a proper set of assessment measures for analyzing intrinsic criminality of communities, especially when considering different crime types. 
    % Results
    We show our methodology by applying it to a real crime dataset from the city of San Francisco -- CA, USA.
    The results confirm its effectiveness to identify and analyze high criminality areas within a city. 
    % Conclusions
    Hence, our contributions provide a basis for further developments on complex networks applied to crime analysis.
\end{abstract}

\IEEEpeerreviewmaketitle

\section{Introduction}
\label{sec:introduction}

Complex networks have long been used to model social behavior, spatial patterns, spreading of epidemics, and urban structures.
They are capable of representing neural connections, data transmission schemes, electric power grids, airports, as well as street, rail and subway networks~\cite{boccaletti2006complex}.
In the context of street networks, which are of particular interest for this work, complex networks are able to describe flow relationships and social behavior in an urban zone~\cite{porta2009street}.

As dynamic organisms, cities evolve (shrink or enlarge) with respect to their space and population, causing changes in their structures.
Such structures are of high relevance as they directly influence the flow within the cities and, consequently, the behavior of their citizens.
Nevertheless, violence and social disorder may also be related to the urban structure of a given city.
For instance, low-flow areas tend to be propitious for crime events, since such areas end up being less surveilled when compared to high-flow areas.
For that reason, crime events usually occur within regions that can be detected based on their structural and behavioral properties~\cite{deryol2016crime}.
In this context, it is our contention that, by means of computational techniques, it is possible to identify and understand the crime dynamics in an urban network.
As so, this paper provides a methodology for aiding the analysis of urban criminality --- \textit{e.g.} how specific crimes occur in certain areas and how their types are related to each other by being in neighboring areas, that is, by being similar.

The hypothesis of this work is that \emph{by employing network mapping techniques, allied to distance-based properties of graphs, it is possible to identify and trace the relationship between areas that are highly criminal within a city}.
Our assumption is that similar crimes occur in adjacent regions.
Such crimes emerge from, among other factors, the urban organization, and their inter-similarity carries correspondence to the distance from one region to the other.
Technically, our proposal is based on: (i) community detection algorithms applied to datasets of isolated crime types; (ii) pattern and similarity analyses, which compare crimes occurred in distinct regions within the same city.
Based on those techniques, we propose a threefold methodology consisted of: \textbf{(i) Mapping of Urban Crimes} -- we describe how to combine a city's georeferenced urban structure with crime records into a complex network with potential for analytical tasks; \textbf{(ii) Criminal Community Identification} -- we show how to detect criminal communities by exploring geographical as well as structural properties; \textbf{(iii) Crime Analysis} -- we methodologically compare crime patterns by identifying relationships between their communities, especially in the face the of diverse crime types.

Our methodology contributes to the understanding of criminality in the context of urban organization.
We work on issues related to:
(i) \emph{an approach to identify highly criminal areas};
(ii) \emph{the characterization of the city space by inferring the homogeneity of its crimes};
(iii) \emph{the identification of city sectors that present behaviors related to the sparsity of their crimes}; and
(iv) \emph{the delineation of the city space by identifying regions that are related, acting as crime spots for different crime types}.

The remainder of the paper is structured as follows.
Section~\ref{sec:related_work} presents the related work, discussing how our methodology stands out from theirs.
Section~\ref{sec:methodology} describes the dataset used in this work and the methodology proposed.
Section~\ref{sec:results} discusses the results obtained from the application of our methodology to the crimes dataset, which describes a real scenario from the street network of the North American city of San Francisco.
Finally, Section~\ref{sec:conclusion} presents the conclusion.

\section{Related Work}
\label{sec:related_work}

%The use of complex networks allows obtaining insights regarding the urban planning process of a city, particularly when considering public safety.
%Therefore, it is important to perform compelling crime analyses considering the real-world space.
%This is because the understanding of criminal patterns and their distribution is the beginning to prevent crimes and to contain their dispersion.

%\textcolor{red}{
Several works in the literature have dealt with the topics of urban networks and criminality. 
This section describes some of these works organized into two categories related to the phases of our proposal: \textbf{Mapping of Urban Crimes}, which focuses on representing raw crime data with a complex network, and \textbf{Crime Analysis}, which refers to identifying criminal behaviors and patterns by analyzing urban street networks.
%}

\vspace{.15cm}\noindent{\bf Mapping of Urban Crimes.}
Spicer \textit{et al.}~\cite{Spicer2016} describe a theoretical framework capable of mapping urban crimes into related locations in a city.
Their work discusses the pros and cons of mapping methods.
To represent a city, they use street networks and GIS software, describing methods for mapping georeferenced elements into a complex network.
However, their methods associate a crime with the nearest node or edge based on address geocoding, not having any relation with graph measures.
Moreover, all methods are superficially introduced, not presenting any formal fundaments.

Shinode and Shinode~\cite{Shiode2013} aim at introducing a search-window method to analyze urban crimes in street networks.
They represent a search window as a subarea that concentrates a high number of crimes in small regions; besides that, the crimes are used to identify clusters spatially and temporally.
The authors validate their methods through an empirical case study.
However, their criminal data derive from 911-calls made in 1996; that is, they used a dataset that carries few details and that is outdated, especially when one considers the conclusions claimed by the authors.
Such information implies a superficial analysis of the crime behavior and the ways to prevent it.

Those previous works aim at identifying highly criminal areas to predict future criminal events.
They use several properties of the data to map them into a node, an edge or a combination of both.
However, they use non-graph measures to build a city scenario, which fails in characterizing the real world.
Our approach, on the other hand, benefits from modeling the real-world space via a georeferenced complex network; a better representation of the reality.
It allows us to merge two georeferenced grids, one describing crimes and the other outlining city elements, which enables us to characterize crimes and extract their patterns.

\vspace{.15cm}\noindent{\bf Crime Analysis.}
White \textit{et al.}~\cite{White2015} represent criminal activities via a georeferenced complex network, through which they compare criminal demographics and identify criminal communities.
The demographic regions are created by clustering the urban areas using socioeconomic borders, whereas the crime network is formed by linking crime locations that are close according to the Euclidean distance.
To identify the criminal communities, the authors employ a label propagation technique.
Their results show that the crime network can be constructed without requiring information about individual crimes.
However, their crime network does not take the urban topology into account, so their comparison is based only on socioeconomic information.

Rey \textit{et al.}~\cite{Rey2012} employ spatiotemporal techniques to represent and analyze urban crimes.
Their approach provides a way to quantify neighborhood criminality and is able to identify patterns from the data.
Moreover, they discuss spatial crime analyses as a way to determine the influence of a crime on its surroundings, analyzing an entire city and providing an evaluation of its crimes.
However, their work is based on outdated data, derived from a police district in the period from 2005 to 2009.
Besides, their methods rely on black-box GIS technologies, used to summarize criminal areas by geocoding each crime into a quarter-mile grid cell.
Furthermore, they analyze only city cells having residential units.

Other works include the proposals of Galbrun \textit{et al.}~\cite{Galbrun2016}, who focus on crime mapping to estimate the probability of a crime occurrence within a street segment; of Fitterer \textit{et al.}~\cite{Fitterer2015}, who use a statistical model to predict crimes in the city of Vancouver; and of Bogomolov \textit{et al.}~\cite{Bogomolov2014}, who predict \emph{crime hotspots} in London by using mobile data. 
Our work differs from theirs because (i) we consider crimes within a more complete and up-to-date period; (ii) we map crimes directly, \textit{i.e.} there is no space summarization in the crime mapping phase; and (iii) we benefit from complex networks, which are a reliable toolset to design georeferenced data.

Considering real crime events and a real-world city, like those presented in this section, we introduce ways to map and analyze crimes, supported by criminal communities extracted from complex networks, allowing outcomes far more detailed and advanced than the former works.
We not only show how to map crimes into georeferenced complex networks but also how to analyze their impact on the city, as well as the relationships among different crime types.
Mainly, we describe results from various points of view, for multiple crime types, and based on several interrelated metrics.

\section{Proposed Methodology}
\label{sec:methodology}

This work aims at analyzing the impact of the urban structure on city criminal events, providing insights to enhance the city planning in highly criminal areas.
To achieve that goal, we propose a methodology with three phases:
(i) \emph{Mapping of Urban Crimes}; (ii) \emph{Criminal Community Identification}; and (iii) \emph{Crime Analysis}.
This section describes such phases according to an application example.

\subsection{Background and Datasets}
\label{sec:background}

We refer to an urban street network as a directed georeferenced graph $G=\{V, E\}$
composed of a set $E$ of $|E|$ edges (street segments) and another set $V$ of $|V|$ nodes (streets intersections).
We refer to an edge $e \in E$ as an ordered pair $\left<i,\ j\right>$, $i \in V$ and $j \in V$, in which $i$ is named \emph{source} and $j$ is named \emph{target}.
Each node $v \in V$ has coordinates $l_{at}$ and $l_{on}$, where $l_{at}$ denotes the node's latitude and $l_{on}$ the longitude.
We refer to a set $C$, $C \cap V = \emptyset$, of crimes, each with coordinates $l_{at}$ and $l_{on}$ within the area of the city.
Each crime $c$ has a value that belongs to a domain of 39 crime types --- the most common types are \textit{assault}, \textit{theft}, and \textit{minor crimes}.

\vspace{.15cm}\noindent\textbf{Graph Source.}
Our approach uses electronic cartographic maps extracted from the \emph{OpenStreetMap} (OSM) ~\cite{haklay2008openstreetmap} platform.
OSM provides maps representing spatial elements, \textit{e.g.} points, lines and polygons, as objects.
Such objects, which are abstractions from the real-world geographical space, are represented by a triple of attributes that uniquely identifies them in their set, granting each object an identification number and georeferenced coordinates.
These objects allow to represent OSM data as a georeferenced graph.

\vspace{.15cm}\noindent\textbf{Crime Source.}
The crime dataset $C$ comes from the \emph{San Francisco OpenData} (SFO) initiative~\footnote{~Available on ``data.sfgov.org''.}, a central clearinghouse for public information about the city of San Francisco -- CA, USA.
SFO provides data concerning public issues.
We use a set of events that corresponds to the period between Jan 1st, 2003 through May 5th, 2016.
It has 1,916,911 instances of 39 crime types.

\subsection{Mapping of Urban Crimes}
\label{subsec:mapping_urban_crimes}

The first phase of our methodology receives the crime dataset as input, denoted as set $C$, where each element $c \in C$ corresponds to a crime event occurred in a specific latitude and longitude.
We proceed by mapping the crime events to the nodes of the urban street network, represented as the graph $G = \{V,E\}$.
Given a crime $c$, its corresponding network node is the closest node $v$, $v \in V$, as given by the Euclidean distance calculated from the coordinates of $c$ and $v$.
This distance refers to the real length between elements thought the Earth's surface, which is derived from the law of cosines and defined as:
\begin{align}
    d_{ij}^{E} = \mathcal{R} \times cos^{-1}({\ }&sin(l_{at}^i)sin(l_{at}^j){\ }+ \nonumber \\ &cos(l_{at}^i)cos(l_{at}^j)cos({\bigtriangleup^{l_{on}}_{ij}}){\ })
    \label{eq:SphericalLaw}
\end{align}
where $l_{at}^i$ and $l_{at}^j$ are latitudes, $\bigtriangleup^{l_{on}}_{ij}$ is the difference between longitudes $l_{on}^i$ and $l_{on}^j$, $\mathcal{R}$ is the earth's radius (6,371km), and $d_{ij}^{E}$ denotes the Euclidean distance ($E$) between points $i$ and $j$.
All values are represented in radians~\cite{wylie2011introduction}.
With respect to the georreferenced network, $d_{ij}^{E}$ also represents an edge's weight, which refers to the street length (in meters) that connects the streets' intersections (\textit{i.e.} nodes) $i$ and $j$.

Next, we enrich the network representation with a new property, making it reflect the geographical distribution of crimes.
The result is a new set of nodes $V'$, with each node $v \in V'$ having a set with three properties $prop(v) = \{l_{at},\ l_{on}, |C_v|\}$, where $l_{at}$ is the latitude, $l_{on}$ is the longitude, and $|C_v|$ is the quantity of crimes mapped to node $v$.
Formally, the set of crimes mapped to a node $v$, denoted as $C_v$, is defined as:
\begin{align}
C_v = \{c \in C ~ | ~ d_{vc}^{E} < d_{wc}^{E}, \forall ~ w \in ~ V, ~ v \neq w\}
\label{eq:Knn}
\end{align}
Note that this phase is performed separately for each kind of crime.
For example, if performing the mapping for crime type \textit{assault}, we get a version of the graph in which the nodes have a property indicating the number of crimes that are categorized as an \textit{assault}.
After processing each kind of crime, we get a new property in the nodes.
In a mapping perspective, each processing yields a new graph that differs with respect to the properties of the nodes; formally,\linebreak $\mathfrak{G} = \{G_0, G_1, \ldots, G_n\}$, where each G = $\{V',E\}, ~ \forall ~ G \in \mathfrak{G}$.

\subsection{Criminal Community Identification}

Intuitively, a community is understood as a group of nodes that have a probability of connecting to each other that is greater than the probability of connecting to nodes out of the group~\cite{Barabasi2014}.
Formally, consider $G=\{V',E\}$ as a city graph and $p(i,j)$ as the probability of nodes $i$ and $j$ to define a connection; a community set is understood as a subset of nodes $\mathfrak{C}$.
A node $i$ is considered a member of the community $\mathfrak{c} \in \mathfrak{C}$ if its probability of connecting to a node $j \in \mathfrak{c}$ is greater than the probability of connecting to a node $j'$ in another community $\mathfrak{c}' \in \mathfrak{C}$, as follows:
\begin{equation}
{p(i,j) \geq p(i,j') ~ | ~ i \not\in \mathfrak{c}, i \not\in \mathfrak{c}', ~ \forall ~ j \in \mathfrak{c}, ~ \forall ~ j' \in \mathfrak{c}', \mathfrak{c} \neq \mathfrak{c}'} \nonumber
\label{eq:community}
\end{equation}
The elements of a community define a connected induced subgraph $G(\mathfrak{c})$ in which each node is reachable from all other nodes of the same community.
Formally, $G(\mathfrak{c}) = \{V'',E'\}$, $V'' \subseteq V' \iff \{\forall ~ i \in \mathfrak{c}, ~ \forall ~ j \in \mathfrak{c} ~ | ~ d_{ij} < \infty\}$,
where $d_{ij}:V' \times V' \rightarrow \mathbb{R}$ is an arbitrary distance function which returns the distance from any given pair of nodes.

Next, we proceed with the identification of communities.
To identify them concerning each crime type, we opted for the tool Nerstrand~\footnote{~Available on ``www-users.cs.umn.edu/~lasalle/nerstrand/''.}, which carries out a fast multi-thread detection.
This tool, which yields results broadly accepted in the literature, identifies communities without any information about how many of them exist in the city.
The input for this phase is an undirected graph, whereas the output is a set of number-labeled communities.
The tool allows the user to attach weights to the graph elements; for our application, the edges' weights are the distance between two distinct nodes, and nodes' properties are the number of mapped crimes.

Our dataset comprises 39 crime types, however, processing and analyzing all of them would exceed the available space of this work.
Besides, our techniques and results are generalizable for any number of crime types.
For those reasons, we worked only with the three most recurrent ones, which comprise almost 40\% of our dataset and represent:
(i) assault, \textit{i.e.} to inflict injury on a person intentionally;
(ii) theft, \textit{i.e.} to take a property off a person's possession by stealth, with no brute force; and
(iii) minor crimes, such as offenses that do not involve any loss or injury.
They represent, respectively, 167,832 (8.76\%), 392,338 (20.47\%) and 204,451 (10.67\%) crimes.
As previously explained, for each crime type, we perform a mapping over the graph, each one with a different number of crimes mapped into their nodes.
Further on, we discuss that each crime-mapped graph can produce a specific criminality scenario; that is, when considering each crime type, we get a different set of communities, each one characterized by that kind of crime and by the topology of the network.

\begin{figure}[!htb]
    \center
    \includegraphics[width=\linewidth]{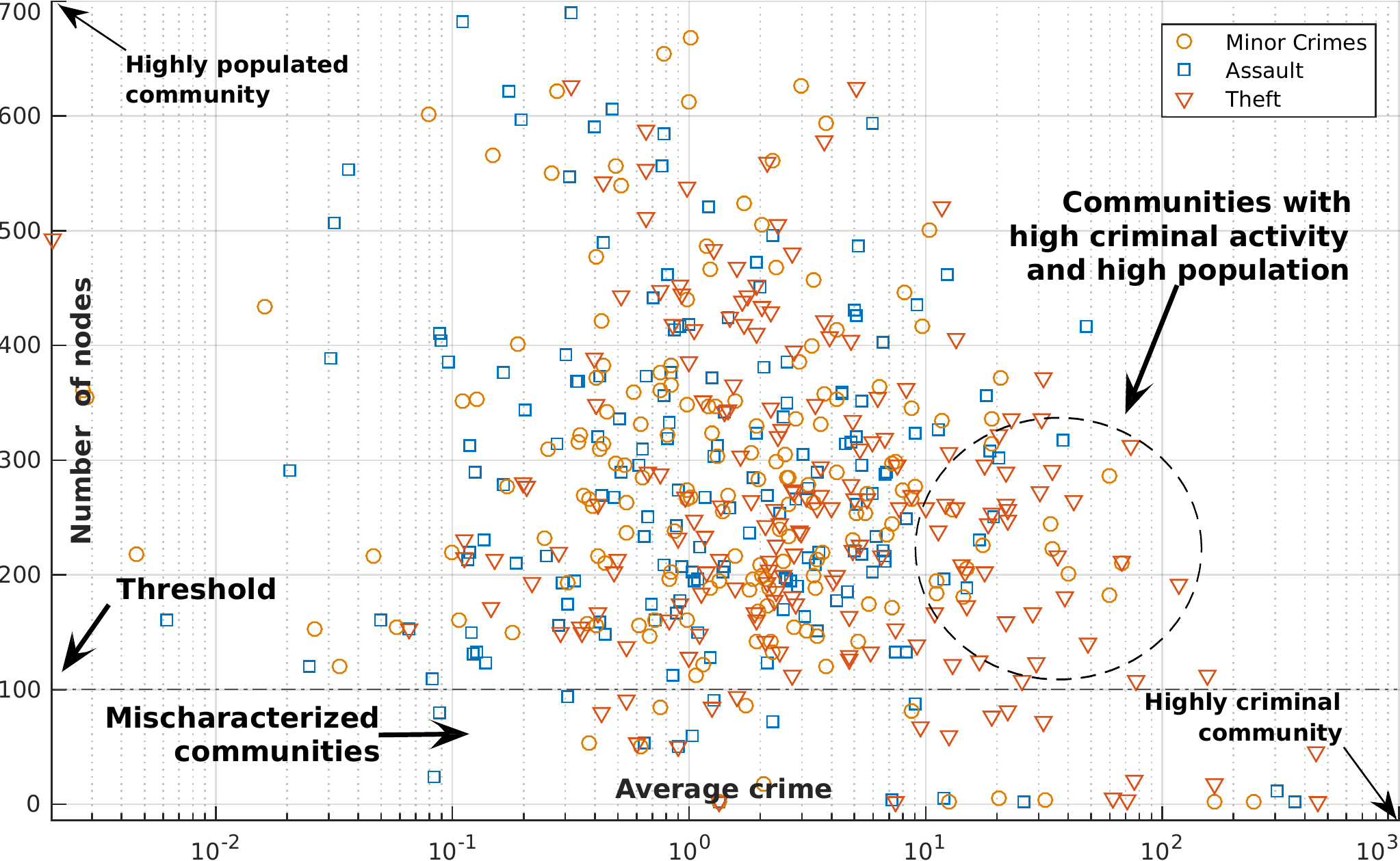}
    \caption
    {Filtering process used to identify the relevant communities while discarding the mischaracterized ones.
    We sort the communities by their average criminality, and we eliminate the spatially-small ones (having less than 100 nodes).
    The highlighted area indicates the communities that are, at the same time, highly populated and highly criminal.}
    \label{fig:CommunitiesInfo}
\end{figure}

The next step, after identifying communities, is to filter out the less relevant ones for each type of crime.
From the perspective of criminality analysis, relevance here means choosing the highly criminal communities.
This approach reduces the excessive number of identified communities, as well as characterizes the most relevant ones.
Accordingly, we worked with the \emph{top five highly criminal communities}, which were identified by using two variables:
(i) the highest criminal average, \textit{i.e.} the average number of crimes per node;
(ii) community size threshold, which prioritizes communities containing at least a specified number of nodes.
For the purpose of this work, the threshold has been set to 100 nodes, since communities with at least 100 nodes are spatially well distributed, and can better represent the regions of the city.
This filtering process is depicted in Figure~\ref{fig:CommunitiesInfo}, which shows the average criminality on the $x$-axis and the number of nodes on the $y$-axis.
In the figure, we describe the less relevant communities as the mischaracterized ones, and we show where the most relevant ones are, considering the previously defined threshold.

\subsection{Crime Analysis}
\label{subsec:AnalyzingCrimes}

In this last phase, with the communities already detected, we proceed to analyze their interrelationship and interaction.
Therefore, here, we consider the communities in pairs of types; for example, we analyze communities of assault and theft to verify whether they occur together or if one type of crime makes the other more intense.
The idea is to obtain insights about their geospatial dependency, pointing to regions that share space of more than one crime type and, consequently, are potentially more dangerous.
To this end, this phase has two stages.
The first stage obtains the geospatial similarity between pairs of crime types, which allows determining how close are their communities and how related are their criminal activity.
The second stage computes the Homogeneity and Completeness scores, which are useful to understand the crimes' behaviors by evaluating specific aspects of each community set.
Combined, both stages lead to a more meaningful analytical scenario, accomplishing the contributions mentioned at the end of Section~\ref{sec:introduction}.

\vspace{.15cm}\centerline{\textbf{Measuring the Similarity of Communities}}\noindent
A similarity value enables us to characterize the relationship of distinct events based on their spatial behavior, more specifically their spatial distribution.
Such similarity is based on their distances and considers the possibility of a non-criminal region containing crimes as well.
This is because a non-criminal area may have its behavior affected by a close criminal area.

We proceed with the computation of the similarity measure by considering the previously detected top five highly criminal communities.
The similarity of crime events is based on the distance between elements belonging to any two sets of communities $\mathfrak{E}$ and $\mathfrak{F}$, where each one represents a distinct crime type.
Note that, at this point, after communities were detected based on the topology of the city, we proceed using only the nodes and their positioning --- \textit{i.e.} we do not consider the edges nor the city topology for this processing.
Such additional information is not necessary since we consider only the position of highly criminal communities in order to characterize the geospatial relationship of crime types.
Hence, we define the similarity $\mathcal{S_K}(\mathfrak{E},\mathfrak{F}): \mathfrak{C} ~ x ~ \mathfrak{C} \rightarrow [0,1]$; the closer to 0, the more spatially apart are the sets of nodes:
\begin{align}
    \mathcal{S_K}(\mathfrak{E},\mathfrak{F}) = 1 - \frac{\sum_{}^{\forall u \in \mathfrak{E}} ~ \sum_{}^{\forall v \in \mathfrak{F}} d_{uv}^{E}}{|\mathfrak{E}| + |\mathfrak{F}|}
    \label{eq:distCluster}
\end{align}
Such measure identifies, at the same time, the similarity between intra-and inter-community elements.
The result $1$ denotes that the elements within community $\mathfrak{E}$ are highly close to each other and to the elements of community $\mathfrak{F}$, intra-and inter-similar respectively, and $0$ denotes the opposite.

\vspace{.15cm}\centerline{\textbf{Identifying the Behavior of Crimes}}\noindent
The characterization of communities identified in a city can be complemented by the Homogeneity and Completeness scores.
They are derived from an entropy-based cluster evaluation~\cite{rosenberg2007}, which works by analyzing the presence and absence of criminality in the network's communities.
Such scores quantify how intensely each crime type manifests across a community set $\mathfrak{C}$.
To this end, the Homogeneity score \emph{evaluates the criminality} in a community by identifying the criminal distribution among their nodes.
Further, the Completeness score is used to \emph{quantify how intrinsic is the criminality} in a community by measuring their distribution among all the communities, both in terms of a specific crime type.

\noindent\textbf{Homogeneity.}
Regarding the top-five criminal communities, it is desired to know what are their predominant aspects.
For the purpose of achieving a better assessment of criminal communities, the Homogeneity score is able to measure how uniform, \textit{i.e.} homogeneous, a community is.
It is defined by $\mathcal{H}(\mathfrak{C}): \mathfrak{C} ~ x ~ \mathfrak{C} \rightarrow [0,1]$ as follows:
\begin{align}
    \mathcal{H}(\mathfrak{C}) =
    1 -
        \frac{
            \sum_{i=1}^{|\mathfrak{C}|} \sum_{j=0}^{|Q|} \frac{|\mathfrak{C}_{i_j}|}{|V'|} \times log_2 \left( \frac{|\mathfrak{C}_{i_j}|}{|\mathfrak{C}_i|} \right)
        }{
            \sum_{k=0}^{|Q|} \frac{|\mathfrak{C}_k|}{|V'|} \times log_2 \left( \frac{|\mathfrak{C}_k|}{|V'|} \right)
        }
    \label{eq:Homogeneity}
\end{align}
\noindent
where $|V'|$ is the number of nodes in all the communities of set $\mathfrak{C}$ (the top five in our example) of a crime type, and $Q = \{1,0\}$, which, in our scenario, indicates the nodes with and without crimes, respectively.
% Even if calculated separately, both scenarios need to be considered to measure the Homogeneity score of a community set.
We used $\mathfrak{C}$ to denote the set that contains all communities and $\mathfrak{C_i}  \in \mathfrak{C}$ to indicate a single community.
Finally, $\mathfrak{C_{i_{j}}}$, $j \in Q$, represents nodes in community $i$ in which crimes are present or absent, whereas $\mathfrak{C}_k$, $k \in Q$, represents the same idea but among all communities.
For instance, $\mathfrak{C}_k$, $k=0$, represents all nodes in all communities that have no crime, while $\mathfrak{C_{i_{j}}}$, $i=1$, $j=1$, indicates the nodes from community 1 that have crime(s).

The Homogeneity score equals 0 when the presence and absence of crimes are proportionally similar across all the communities.
Contrarily, the value 1 denotes that all communities in $\mathfrak{C}$ have crimes in all their nodes or no crime at all.

It is noteworthy that the Homogeneity score solely is insufficient to provide the needed comprehension about the criminal behavior in a city.
This is because a single community can have all its nodes characterizing either crime occurrences or no crime at all, which, in turn, could yield the same Homogeneity score.
For this reason, we also employ the Completeness score that, combined with the Homogeneity score, can better describe the distribution of crimes across a city.
By analyzing both scores, we can effectively compare different areas, regardless of their high or low criminality.
Their purpose is to identify the communities' behaviors, mainly their tendency for high/low scores to a given crime type.

\vspace{.15cm}\noindent\textbf{Completeness.}
If crimes do not concentrate in a community, it is likely that there are multiple communities with criminal nodes spread across the city, demanding surveillance at multiple criminal spots.
Alternatively, if crimes concentrate, they tend to be grouped within a single community, requiring police patrols on specific regions.
To determine if crimes are more or less concentrated, we employ the Completeness score, which is directly obtained from the whole community set.
The Completeness score equals 0 when the presence of crimes is totally scattered, \textit{i.e.} equally absent across all communities --- their occurrence is totally distributed.
On the other hand, the higher the value, the more concentrated are the criminal and safe zones.
This is defined in $\mathcal{C}(\mathfrak{C}): \mathfrak{C} ~ x ~ \mathfrak{C} \rightarrow [0,1]$ as:
\begin{align}
    \mathcal{C}(\mathfrak{C}) =
    1 -
        \frac{
            \sum_{j=0}^{|Q|} \sum_{i=1}^{|\mathfrak{C}|} \frac{|\mathfrak{C}_{i_j}|}{|V'|} \times log_2 \left( \frac{|\mathfrak{C}_{i_j}|}{|\mathfrak{C}_j|} \right)
        }{
            \sum_{k=1}^{|\mathfrak{C}|} \frac{|\mathfrak{C}_k|}{|V'|} \times log_2 \left( \frac{|\mathfrak{C}_k|}{|V'|} \right)
        }
    \label{eq:Completeness}
\end{align}
Hence, if safe and criminal zones are totally apart, the Completeness score of a community set equals 1.
Conversely, if criminality has a proportionally equivalent participation in all communities, the score is 0.

\section{Results and Discussions}
\label{sec:results}

This section presents and discusses the results of the experiments regarding the three phases of our methodology. % over the city of San Francisco.

\begin{figure*}[!htb]
    \centering
    \includegraphics[width=\linewidth]{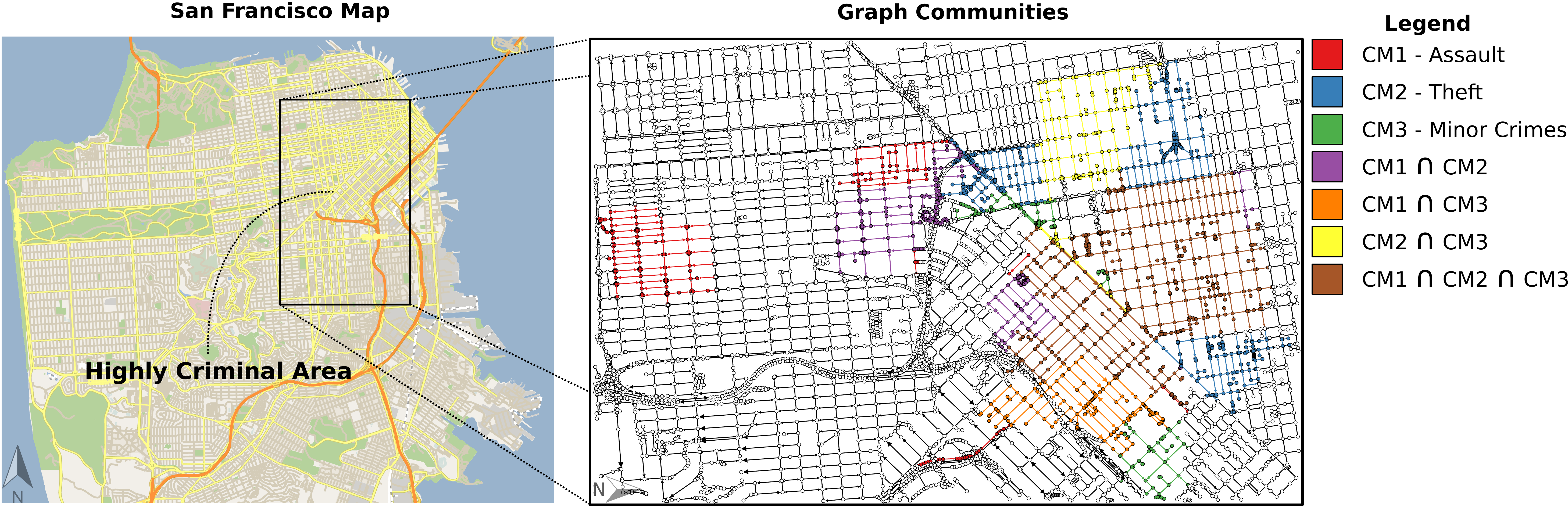}
    \caption{The most highly criminal region from San Francisco -- CA, USA, identified by evaluating the crime density in the city from 01/01/2003 to 05/18/2016.
    Each colored area represents the occurrence of a crime type or the intersection of a crime type with another type of crime.
    In the legend, each CM (from 1 to 3) denotes a type --- \textit{assault}, \textit{theft} and \textit{minor crimes}.}
    \label{fig:SanFran}
\end{figure*}

\vspace{.15cm}\noindent\textbf{Influence of Crime Mapping.}
The first experiment aimed at analyzing the impact of the urban crime mapping (first phase of our methodology) on the criminal community identification process.
This process was carried out twice: once considering the topological graph, \textit{i.e.} the graph that considers only nodes and edges, and once over the network after the urban crime mapping, regarding the newly-included edges' weights (distances in meters) and nodes' weights (number of crimes).
As a result, we obtained the number of 134 communities from the topological graph and, when considering the information introduced by the crime mapping phase, the numbers of identified communities were:
210 for \textit{assault},
215 for \textit{theft} and
211 for \textit{minor crimes}.

As a first result, we noticed the discrepancy between the numbers of criminal communities for each graph.
Such difference is due to the presence/absence of crime-related data derived from the crime mapping phase.
The Nerstrand algorithm considers the crime-related data (the nodes' and edges' weights), ensuring that the nodes within each identified community share similar characteristics.
This behavior is not the same when considering the topological graph.
Since there is no crime information attached to its nodes and edges, the algorithm examines only the connections between nodes.
Therefore, considering only the network's topology, there is no guarantee that the crimes will be optimally grouped among the identified communities.
These findings show that identifying communities through a topological graph is an inaccurate way to represent crime relationships in a city.

\vspace{.15cm}\noindent\textbf{Communities' Design Aspects.}
The second experiment aimed at comparing the results from the previous analysis by considering their design aspects.
Table~\ref{table:top12} presents the most highly criminal communities obtained directly from the network's topology and from the three graphs generated via the Crime Mapping process.
The table shows twelve communities sorted by crime average, among which are the top five criminal ones.% for each type.

\begin{table}[!htb]
    \centering
    \caption
    {Communities with the highest crime average, identified from: (i) \textit{topology}, (ii) \textit{assaults}, (iii) \textit{theft} and (iv) \textit{minor crimes}.
        Column \textit{Avg} denotes the crime average and \textit{\#} denotes the number of nodes in each community.}
    \scalebox{.92}{
        \begin{tabular}{c|c|c|c|c|c|c|c|c}\hline
            \multirow{3}{*}{}  & \multicolumn{8}{c}{\textbf{Communities identified from:}} \\ \hline
            \rowcolor{gray!20} & \multicolumn{2}{c|}{\textit{Topology}} & \multicolumn{2}{c|}{\textit{Assault}} & \multicolumn{2}{c|}{\textit{Theft}} & \multicolumn{2}{c}{\textit{Minor Crimes}} \\ \hline
            & Avg & \# & Avg & \# & Avg & \# & Avg & \# \\ \hline
            \rowcolor{gray!20}00 & 1063.00 & 2 & 362.00 & 2 & 4030.00 & 3 & 545.92 & 12 \\ \hline
            01 & 157.67 & 788 & 230.59 & 17 & 245.00 & 2 & 457.00 & 2 \\ \hline
            \rowcolor{gray!20}02 & 116.30 & 802 & 55.26 & 316 & 201.00 & 17 & 356.41 & 17 \\ \hline
            03 & 110.75 & 4 & 47.44 & 226 & 60.15 & 281 & 281.91 & 67 \\ \hline
            \rowcolor{gray!20}04 & 108.80 & 5 & 41.00 & 2 & 49.74 & 277 & 140.13 & 142 \\ \hline
            05 & 77.93 & 515 & 34.69 & 87 & 47.08 & 156 & 110.15 & 188 \\ \hline
            \rowcolor{gray!20}06 & 77.50 & 2 & 34.06 & 149 & 37.88 & 66 & 73.80 & 5 \\ \hline
            07 & 65.86 & 539 & 29.53 & 238 & 32.25 & 4 & 72.99 & 143 \\ \hline
            \rowcolor{gray!20}08 & 44.17 & 744 & 27.42 & 142 & 25.92 & 296 & 71.37 & 286 \\ \hline
            09 & 43.88 & 1129 & 22.14 & 248 & 25.77 & 342 & 68.50 & 4 \\ \hline
            \rowcolor{gray!20}10 & 36.28 & 635 & 15.52 & 281 & 21.79 & 307 & 58.72 & 32 \\ \hline
            11 & 35.88 & 518 & 14.20 & 54 & 21.10 & 371 & 54.43 & 144 \\ \hline
        \end{tabular}
    }
    \label{table:top12}
%    \vspace{-.5em}
\end{table}

The table describes a higher crime average in communities derived from topology only.
Such higher average occurs because, in this context, there is a lower number of communities, each one containing a higher number of criminal nodes.

As shown in the table, the number of crimes per community still stands out when comparing the graph from topology and the ``complete'' graphs --- \textit{i.e.} the graphs generated by the Crime Mapping process.
This is because the complete graph used no attribute that characterizes a crime in the community identification process, so there is no way to assure that they are related just by being in the same community.
Therefore, hereafter we discuss only the graphs with crime-related data.

Regarding such graphs, Table~\ref{table:top12} shows a significant number of communities with few nodes.
This is evidence that the crimes are not evenly distributed in the network.
% Therefore, if someone eventually attempted to eradicate or prevent crimes in the entire city based on characteristics of a local region (a single community), such attempt would fail
Therefore, if someone eventually attempted to eradicate or prevent crimes in the entire city based on characteristics of a local region, such attempt would fail because the crimes are not concentrated in a single criminal community.

\vspace{.15cm}\noindent\textbf{San Francisco's Crime Neighborhood.}
To allow analyzing a group of communities containing a significant number of crimes, we have selected the region from the city of San Francisco that comprises the highest crime indices.
This region contains the five communities, as mentioned earlier, of each crime type, which are among the twelve communities shown in Table~\ref{table:top12}.
The top five communities of each type also meet the requirement regarding the number of nodes, previously determined as a threshold of 100 nodes (see Section~\ref{subsec:AnalyzingCrimes}).
Specifically, considering the table, the top five communities for \textit{assault} are in lines 2, 3, 6, 7 and 8; the top five for \textit{theft} are in lines 3, 4, 5, 8 and 9; and the top five for \textit{minor-crimes} are in lines 4, 5, 7, 8 and 11.
Figure~\ref{fig:SanFran} depicts the entire selected area of San Francisco, as well as highlights each community (and each geospatial intersection of communities).% with respect to their crime types.

To quantify the relationship between communities in the city space, we used Equation~\ref{eq:distCluster} to obtain the similarity between communities.
Through this measure, we compared each pair of crime types, computing the similarity between community sets $\mathfrak{E}$ of one type and community sets $\mathfrak{F}$ of another type.
By doing so, we achieved a similarity value $\mathcal{S_K}(\mathfrak{E},\mathfrak{F})$ of $0.66$ for assault~\textit{vs.}~theft, $0.65$ for assault~\textit{vs.}~minor crimes and $0.58$ for theft~\textit{vs.}~minor crimes.
Such results suggest that there is a relationship between crimes of different types, especially when comparing assault~\textit{vs.}~theft and assault~\textit{vs.}~minor crimes, since the values $0.65$ and $0.66$ can be considered significant enough if one ponders over what these numbers represent --- that the majority of the area of these two communities might be more dangerous due to the occurrence of more than one crime type.

%\begin{enumerate}
%\itemsep 0 em
%\item It is very likely that criminal activities occur at intersections of multiple crime types, which we call \textit{crime hubs} --- regions that attract or disseminate criminality;
%\item Those crimes are spatially related, so it is possible that they have been caused by the same people or gang;
%\item Places where more than one crime type occurs are high-priority regions for public safety improvement.
%\end{enumerate}

More specifically:
\textbf{(i)} it is very likely that criminal activities occur at intersections of multiple crime types, which we call \textit{crime hubs} --- regions that attract or disseminate criminality;
\textbf{(ii)} those crimes are spatially related, so it is possible that they have been caused by the same people or gang;
\textbf{(iii)} places, where more than one crime type occurs, are high-priority regions for public safety improvement.

\vspace{.15cm}\noindent\textbf{Communities' Intrinsic Criminality.}
The previous section discussed the geospatial disposition of different crime types with respect to intersections between criminal regions.
This section, in turn, brings results concerning the Homogeneity and Completeness scores for the communities in the most highly criminal area of San Francisco.
This experiment aimed at measuring how intrinsic was the criminality within the neighborhoods of San Francisco during the studied period.
To understand this measure, recall that the number of crimes of some nodes is zero.
For the \textit{assault} type, there are 1,071 nodes, of which 132 (12.33\%) contain at least one crime and 939 (87.68\%) contain none.
For \textit{theft}, there are 903 nodes, of which 133 (14.73\%) are criminal and 770 (85.27\%) are not criminal.
For \textit{minor crimes}, there are 1,352 nodes, of which 177 (13.09\%) present crimes and 1,175 (86.91\%) do not.
These values and percentages refer to the entire community set for each crime type.

The presence and absence of crimes were considered to calculate the scores presented in Table~\ref{table:HC}, which shows that \textit{assault} has the lowest values for both measures.
Regarding Homogeneity, the value indicates that this crime type presents a reduced concentration of criminal nodes within each community.
This is straightly inferable because, besides the low Homogeneity value indicating an inexpressive occurrence of crimes, the actual number of crimes is very low; a pattern that holds for other types of crime as well.

\begin{table}[!htb]
    \centering
    \caption{The Homogeneity and Completeness scores used to evaluate the quality of each community set and to quantify how intrinsic is the criminality within them.}
%    \scalebox{.84}{
\begin{tabular}{r|c|c} \hline
    \textbf{Crime type} & Homogeneity & Completeness  \\ \hline
    \rowcolor{gray!20}  \textit{Theft} & 0.015588 & 0.004159 \\ \hline
    \textit{Minor Crimes} & 0.014014 & 0.003441 \\ \hline
    \rowcolor{gray!20}  \textit{Assault} & 0.013557 & 0.003234 \\ \hline
\end{tabular}
%}
\label{table:HC}
%\vspace{-1em}
\end{table}

Since the value for Completeness is also the closest to 0, we can reliably assume the occurrence of crimes to be scattered across all the \textit{assault} communities.
The \textit{theft} type has values close to 0 as well, but the highest ones when compared to the values of other crime types.
The greater difference corresponded to the Completeness score, whose value suggests a geospatial crime concentration more elevated.
Regarding Completeness (see Section~\ref{subsec:AnalyzingCrimes}), such higher concentration tends to occur in a smaller set of communities rather than across all of them, which is more likely for the \textit{assault} type.

A higher crime concentration means that a particular crime type is less likely to intersect with another crime type.
This is because, in such a case, the criminal nodes would occupy a smaller area in the city space.
Furthermore, such a crime type could be considered easier to prevent, since its occurrences would be in a more compact area of the city.
On the other hand, a lower concentration of a crime type means that it exerts a stronger influence on the network as a whole, being easier to propagate to remote areas and demanding global approaches both to eradicate it and to prevent it.

%\vspace{.15cm}\centerline{\textbf{Remarks}}\noindent
%We focused on analyzing the behavior of criminality in urban areas through complex network tools.
%Due to the lack of space and to the difficulty to acquire well-organized crime data, our work was limited to the city San Francisco, a very representative urban scenario.
%Nevertheless, it is noteworthy that our methodology is adequate for any real-world city, provided the data is available. 
%We demonstrated that the distances between the nodes of a city and the data concerning criminal occurrences can be used in the task of analyzing the criminality of a region, which is done through computational and mathematical methods.
%Nonetheless, we are aware that criminality is a sociology problem to be further analyzed from the perspective of social sciences. This is the reason why we did not attempt to explain the reasons behind the criminal instances; rather, we traced a scenario from the perspective of the relationships between georeferenced interacting crimes. This was also the case of the work of Bogomolov \textit{et al.}~\cite{Bogomolov2014}, which presents the closest methodology as compared to our approach. In their work, the authors do not take social variables into account; still, they are able to predict crime instances from demographic information, despite not explaining the reasons beneath. Complementarily, our work hypothesizes on how crime patterns are geographically related, bringing up another dimension to allow decision making over urban networks.

\section{Conclusion}
\label{sec:conclusion}

In this paper, we propose a threefold methodology for identifying and analyzing criminal spatial patterns in urban street networks.
Our methodology is based on the hypothesis that \emph{by employing network mapping techniques, allied to distance-based properties of graphs, it is possible to identify and trace the relationship between areas that are highly criminal within a city}.
To demonstrate our methodology, we analyzed criminal communities from real crime data representing \emph{assaults}, \emph{thefts} and \emph{minor crimes} in the city of San Francisco -- CA, USA.
The methodology comprises the following phases: (i) Mapping of Urban Crimes, (ii) Criminal Community Identification and (iii) Crime Analysis.
The latter employs the well-established Homogeneity and Completeness scores to analyze the identified criminal communities more deeply.
Our main achievement confirm the hypothesis of our work, allowing us to state that different crime types share common spaces, characterizing areas that lack strategies for crime prevention, and that particular crime types are sparser than others in the city space.

The highlighted contributions are:
\textbf{(i)} the use of a complex network to represent the real-world space, enabling a complete analysis of the city;
\textbf{(ii)} independence from socioeconomic information, allowing the analyses of crimes based on solely their spatial disposition;
\textbf{(iii)} the assessment of the impact of criminal regions, considering the similarity between distinct crime types and their Homogeneity and Completeness scores.

Finally, as future work, we intend to continue expanding our methodology developing tools to analyze crimes and their types considering a temporal perspective.
Our aim is to provide insights about the criminal behaviors in a city, allowing to determine how a crime spreads throughout the city and when interventions were made by the governors were effective to prevent them, using time windows and comparing the results with socioeconomic data.

\vspace{.15cm}\noindent\textbf{Acknowledgments.}
We are grateful to CNPq (National Counsel of Technological and Scientific Development) by the support (grants 9254-601/M, 444985/2014-0, and 147098/2016-5), to FAPESP (Sao Paulo Research Foundation) by the assistance (grant 2015/15392-7, 2016/02557-0, 2016/17330-1), to Capes (Brazilian Federal Agency for Support and Evaluation of Graduate Education) and to the Databases and Images Group (GBdI) by the resources ceded to this research.

\bibliographystyle{IEEEtran}
\bibliography{references}

\end{document}